# A Platform for Far-Infrared Spectroscopy of Quantum Materials at Millikelvin Temperatures

Michael Onyszczak[1,#], Ayelet J. Uzan[1,#], Yue Tang[1,#], Pengjie Wang[1,#], Yanyu Jia[1], Guo Yu[1,2], Tiancheng Song[1], Ratnadwip Singha[3], Jason F. Khoury[3], Leslie M. Schoop[3], Sanfeng Wu[1,*]

[1]Department of Physics, Princeton University, Princeton, New Jersey 08544, USA
[2]Department of Electrical and Computer Engineering, Princeton University, Princeton, New Jersey 08544, USA
[3]Department of Chemistry, Princeton University, Princeton, New Jersey 08544, USA.
[#]These authors contributed equally to this work
[*] Email: sanfengw@princeton.edu

**Abstract:**

Optical spectroscopy of quantum materials at ultralow temperatures is rarely explored, yet it may provide critical characterizations of quantum phases not possible using other approaches. We describe the development of a novel experimental platform that enables optical spectroscopic studies, together with standard electronic transport, of materials at millikelvin temperatures inside a dilution refrigerator. The instrument is capable of measuring both bulk crystals and micron-sized two-dimensional van der Waals materials and devices. We demonstrate its performance by implementing photocurrent-based Fourier transform infrared spectroscopy on a monolayer $WTe_2$ device and a multilayer $1T\text{-}TaS_2$ crystal, with a spectral range available from the near-infrared to the terahertz regime and in magnetic fields up to 5 T. In the far-infrared regime, we achieve spectroscopic measurements at a base temperature as low as ~ 43 mK and a sample electron temperature of ~ 450 mK. Possible experiments and potential future upgrades of this versatile instrumental platform are envisioned.

## I. Introduction

At ultralow temperatures (ULT, < 1 K), quantum materials may develop novel quantum phases, often associated with a small energy gap in the far-infrared (FIR)/terahertz (THz) regime (~ meV energy scale). Some known examples are energy gaps in Landau quantized systems, many-body gaps in correlated or twisted 2D materials, and superconducting gaps/pseudogaps in unconventional superconductors. In this regime, mature experimental characterizations so far are largely limited to charge-based probes, e.g., electronic transport and tunneling techniques, which are sensitive to charged excitations but not charge-neutral modes. Optical spectroscopy can uncover critical information inaccessible to other techniques and potentially access charge-neutral modes hidden to charge transport. An outstanding example, in which optical detection has significant advantages, is the detection of low-energy neutral excitations in strongly correlated quantum matter. In the case of fractional quantum Hall effect (FQHE), one such kind of excitation is the charge-neutral magneto-roton mode[1] (with an energy scale of ~ 1 meV), which was first experimentally revealed using a Raman scattering technique[2] at ULT with high-energy incident photons (> 1 eV). Recent theoretical analysis has revealed new aspects about this neutral mode at its long wavelength limit[3,4] and suggested that a polarization-resolved optical measurement could be used to detect them[4–6]. Like the Raman scattering experiment[2], a few previous experiments on condensed matter samples cooled in a dilution refrigerator (DR) also employed high-energy photons near the visible (sometimes through fiber optics)[7,8]. Yet as far as we know, direct optical studies in the FIR/THz regime at ULT remain largely unexplored.

Another interesting topic is to search for FIR sub-gap optical absorption and possible cyclotron resonances inside a correlated insulating gap, whose detection may serve as signatures of an emergent charge-neutral (spinon) Fermi surface[9,10]. Experimental techniques are currently not available for exploring many of these frontiers. It is highly desired to develop optical spectroscopic techniques to directly access low-energy excitations in quantum materials, which requires performing FIR/THz optical spectroscopy at ULT and under high fields.

In addition, 2D quantum materials and van der Waals (vdW) structures are rapidly advancing in the fields of topological, superconducting, and correlated quantum matter. During the past two decades, a wide range of experimental tools for fabricating high-quality 2D structures and devices have been developed, establishing this class of materials as an exceptionally rich and versatile system for discovering new quantum phenomena. FIR/THz optical measurements of 2D materials and devices[11,12] are currently limited to samples at relatively high temperatures (> 2 K) and its applications to the large class of emergent new 2D materials and vdW structures, such as twisted moiré materials and a wide range of air-sensitive 2D materials, are largely unexplored. The lack of optical characterization tools for 2D samples at ULT leaves behind many open questions and unexplored territories, especially regarding strongly correlated quantum matter and their interactions with light.

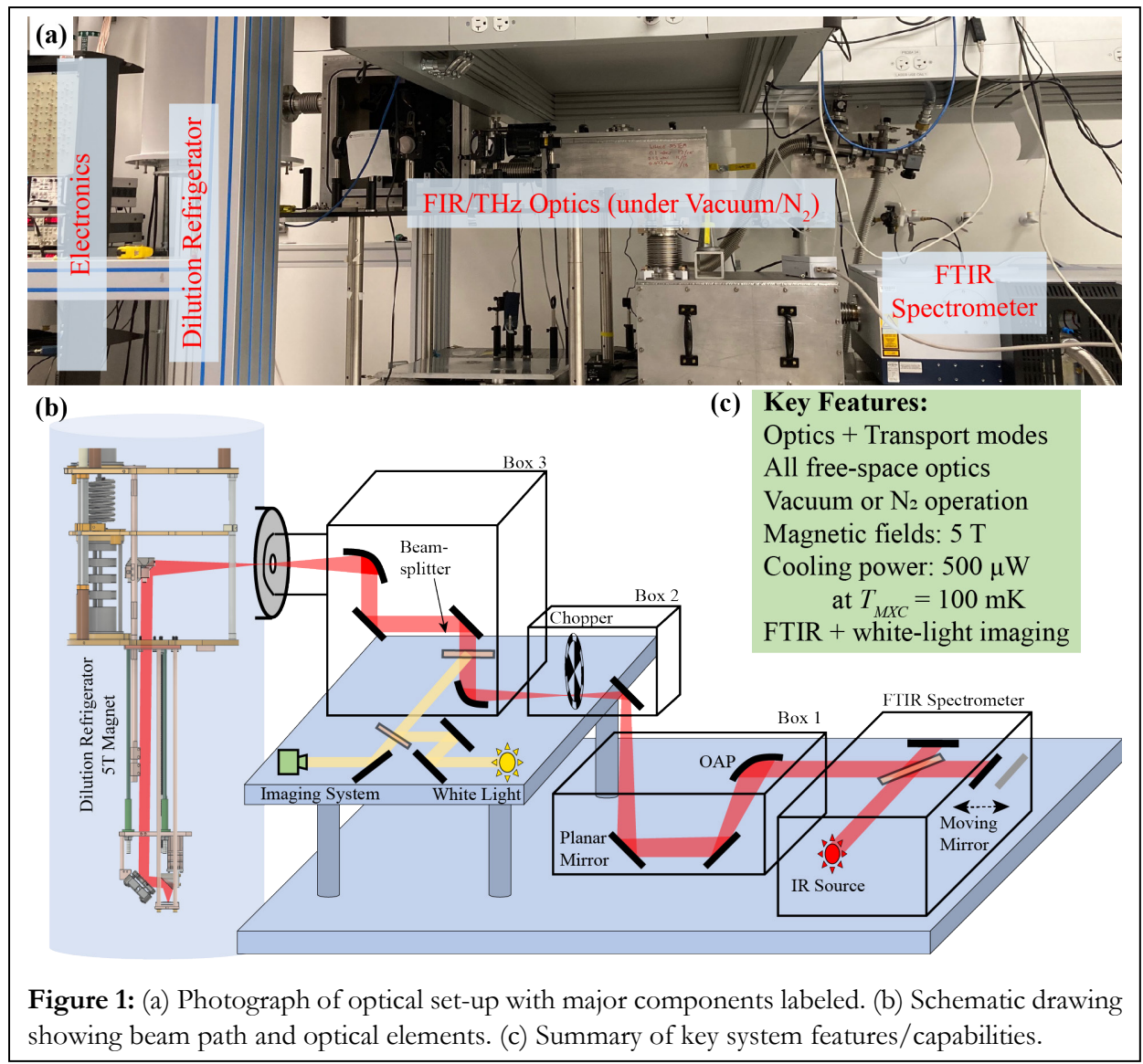

**Figure 1:** (a) Photograph of optical set-up with major components labeled. (b) Schematic drawing showing beam path and optical elements. (c) Summary of key system features/capabilities.

In this work, we report the development of a novel instrument capable of performing all-free-space optical spectroscopy with spectral coverage from the near-infrared to the THz regime, together with standard electronic transport measurements, for characterizing quantum materials at ULT and under magnetic fields. The instrument is designed and built for studying the properties of both bulk crystals and 2D vdW materials and investigating their light-matter interactions.

## II. Experimental Platform

The instrument integrates both optical and electronic probes into a DR system. A photo and a design sketch of the overall setup are shown in Fig. 1a & b. As we elaborate below, the platform overcomes several critical technical challenges, including (1) achieving millikelvin electron temperatures with FIR/THz light on the sample; (2) performing optical spectroscopy at such low temperatures; (3) extracting an optical response from micron-sized 2D samples (in addition to larger bulk crystals) while the optical beam spot is on the millimeter scale; (4) simultaneously performing standard transport measurements on gate-tuned high-quality (fully boron-nitride encapsulated) 2D devices; and (5) integrating magnetic fields. As examples, we demonstrate the implementation of photoconductivity-based Fourier transform infrared spectroscopy (FTIR)[13,14] on both thick crystals and 2D materials. Key features are summarized in Fig. 1c. Below we separately describe the room temperature (outside DR) and low temperature (inside DR) components of the instrument in detail.

### A. Room Temperature External Optics

To enable spectroscopic measurement, we direct the optical beam from an FTIR spectrometer (Bruker VERTEX 80v) onto the sample. We note that while the FTIR experiment is used as a demonstration, with straightforward modifications other types of optical experiments are also possible. In the spectrometer, a wide continuous infrared (IR) source is generated using a globar lamp or a Hg arc lamp and passed through a Michaelson interferometer. The interferometer is equipped with a step-scan mirror mounted on a delay stage which allows the mirror to be moved in repeatable discrete steps with a scan range that yields a spectral resolution of up to ~ 0.025 meV. Using different combinations of the light source, beam splitters, and optical filters (e.g., KBr and Mylar beam splitters), a selected spectral range within the range of ~ 1 meV up to ~ 1 eV is accessible for spectroscopy measurements. In our experiment, the output beam from the interferometer propagates into the DR (Fig. 1b), which places several constraints on designing and building the optics.

To prevent the loss of FIR/THz intensity caused by the strong absorption by atmospheric gases, we place all external optics under vacuum, in three welded aluminum boxes (Box 1-3 in Fig. 1b) capable of being evacuated to a pressure of < 1 mbar (or being purged with nitrogen gas to remove unwanted molecules in the beam path, e.g., $CO_2$, water, etc.). Additionally, we avoid using transmissive optical components for the FIR beam to ensure high and uniform transmission across the entire spectral range, i.e., the FIR beam is directed using gold-coated planar mirrors and focused using gold-coated off-axis parabolic (OAP) mirrors. We have removed the output window of the Bruker spectrometer such that the sealed spectrometer environment is directly connected to Box 1. Namely, the spectrometer experiences the same vacuum or purge conditions as it is directly connected to the boxes. To enter the dilution refrigeration, a diamond window (see details below) is used to allow the light to propagate into the DR while the fridge maintains its own separate vacuum.

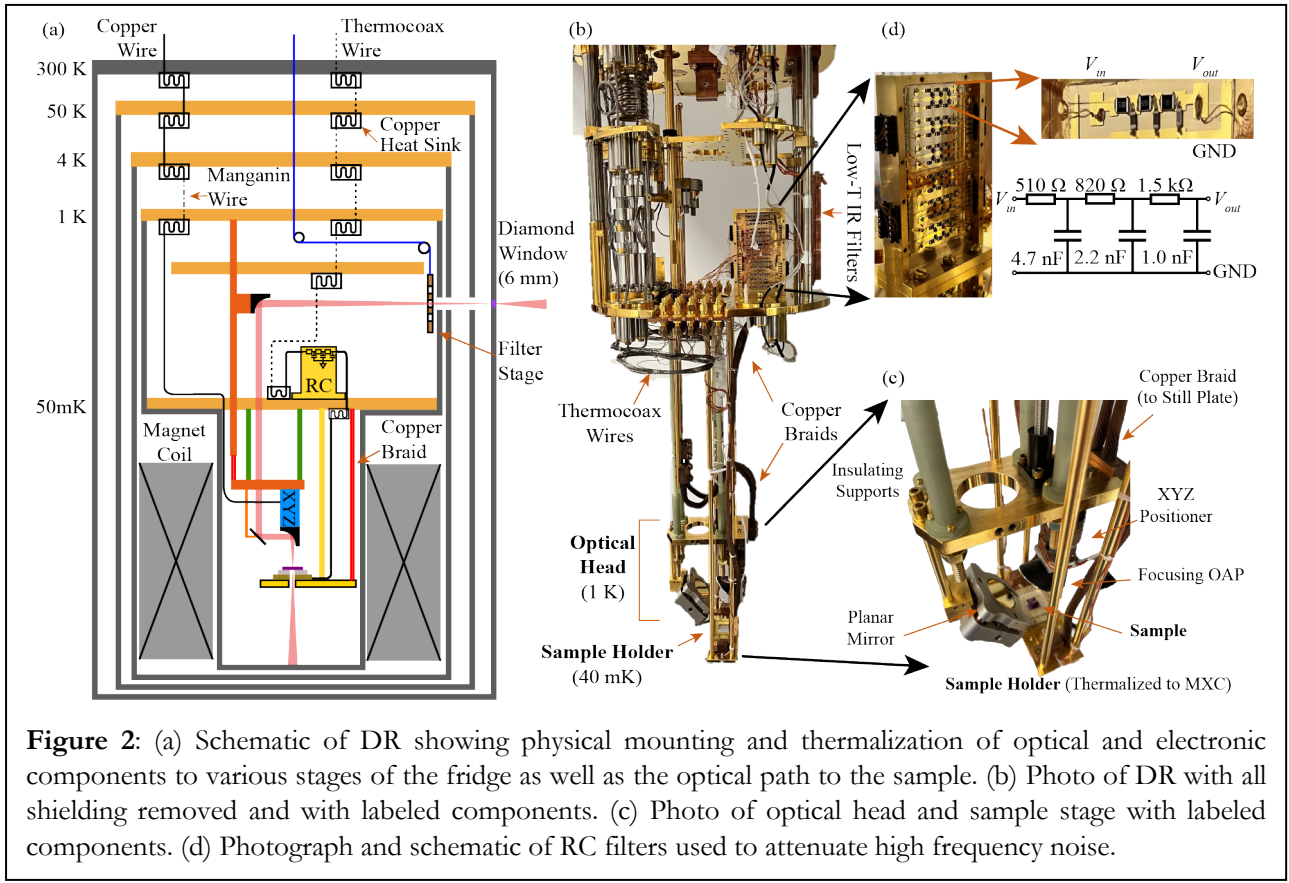

**Figure 2**: (a) Schematic of DR showing physical mounting and thermalization of optical and electronic components to various stages of the fridge as well as the optical path to the sample. (b) Photo of DR with all shielding removed and with labeled components. (c) Photo of optical head and sample stage with labeled components. (d) Photograph and schematic of RC filters used to attenuate high frequency noise.

The spectrometer is placed 1.8 m away from the center of the superconducting magnet (i.e., the sample area) to avoid any effect of the stray field. We place two telescopes, each consisting of two custom OAP mirrors, to direct the beam from the FTIR output into the DR. After the first telescope, the beam size is reduced from ~ 50 mm to ~ 25 mm. At the focus of this first telescope (inside Box 2), a mechanical chopper (Thorlabs MC200B, 4 – 200 Hz) is placed to allow for lock-in measurements. The focus of the second telescope is located at the diamond window of the DR. This minimizes the optical window size on the DR which reduces thermal radiation and maintains the low sample temperature. After the second telescope, the beam inside the fridge has a size of ~ 15 mm, before being focused onto the sample (to be described in the next section).

We have also integrated a white-light imaging system (Fig. 1b) that allows for visualizing the sample inside the fridge, which plays a key role in ensuring that the optical beam is focused onto the 2D samples that are often a few microns in size. In addition, to align the FIR/THz beam, a green laser is used as guidance for the invisible FIR/THz beam (not shown in the sketch). These functions are coupled to the FIR/THz optics using beam splitters, which are removed from the beam path during spectroscopic measurements.

### B. Low-Temperature Optics

The sample and all low-temperature optical components are housed in a DR (Bluefors LD400), as sketched in Fig. 2a. The entrance is a wedged diamond window, which has a 6 mm diameter and 0.5 mm thickness and is mounted in a custom KF-50 flange. The use of a diamond window allows for good transmission for all-optical frequencies ranging from the visible (for white light imaging) to the FIR/THz range. The beam then passes through three, 9 mm, open apertures in the 50 K, 4 K, and 1 K stages, respectively, of the DR before being steered downwards toward the magnetic field center by an OAP mirror. This OAP mirror is the second focusing mirror of the second telescope mentioned above, whose focus lies at the diamond window at the entrance to the DR. Next, the collimated beam is steered using a planar mirror before being focused using a 25 mm focal length gold-coated OAP mirror onto the sample, which is located at the magnetic field center. The final OAP mirror is mounted on an XYZ positioner stack (Attocube ANPx101/ULT/RES+/HV & ANPz102/ULT/RES+/HV) to allow for beam positioning and scanning across the sample. To fulfill the requirements of < 10 Ω wire resistance for the attocube positioners, we split the wiring at the 4K stage into two segments: From room temperature to the 4 K stage, twisted copper wires are used and thermalized at 50 K and 4 K stages; and between the 4 K stage and 1 K stage, we use manganin wires (d = 0.5 mm, California Fine Wire Co.) to provide low resistance and low thermal conductivity. The resistance of a single line from room temperature to the 1 K stage is 5 Ω. The attocube positioners are then connected to the cables using manually twisted pairs of copper wire (d = 0.15 mm, Block Transformatoren, CUL 100/0,15) that are thermalized to the 1 K stage.

To minimize the heat load on the mixing chamber (MXC) plate and to maintain the low base temperature, all optical components (labeled as "optical head" in Fig. 2b & c) are thermally connected to the still plate (1 K) instead of the MXC plate which is at the lowest temperature and provides cooling to the sample itself. The mechanical support of the optical head is provided using 4 insulating G-10 rods connected to the MXC plate, which reduces the heat transfer from the optical head to the MXC plate. To cool the optical head, which includes mirrors and the attocube positioner, a gold-coated, copper rod is connected to the 1 K plate, passes through a hole in the MXC plate, and is thermally connected to the optical head using two welded copper braids.

To reduce unwanted radiation entering the system, e.g., ambient temperature blackbody radiation, a movable optical filter stage is mounted after the 1 K aperture. This cryogenic optical filter system, thermalized at 1 K, consists of a modular copper filter holder that holds up to three 1" diameter filters (or any 1" optical component) that ride along two G-10 guide rails vertically. The filter holder is connected via a Kevlar string and a pair of pulleys to a linear translator mounted via a vacuum feedthrough on the top of the DR (Fig. 2a).

### C. Low-Temperature Electronics

To minimize the sample electronic temperature, we built a wiring system in which thermal links are carefully engineered and high-frequency electromagnetic noises are heavily filtered. To filter the MHz to GHz range, we used Thermocoax cables ($d_{core}$ = 0.5 mm, model # 1Nc Ac o5) which run from room temperature to the mixing chamber and are thermally anchored at the 50 K plate, 4 K plate, 1 K plate, intermediate cold plate (~ 100 mK), and the MXC plate (Fig. 2a & b). At each plate, the cables are wound around copper bobbins filled with silver paint (SPI Supplies, 05002-AB). The bobbins are shielded with aluminum cylinders and are anchored at each plate. At the MXC, the Thermocoax cables are connected to a set of low-temperature RC filters installed on a pair of custom gold-coated copper stages. Each filter stage holds 24 lines mounted on sapphire substrates (Fig. 2d), which are designed to filter out frequencies in the kHz to MHz range. The output of the RC filters connects to manually twisted pairs of copper wires (d = 0.15 mm, Block Transformatoren, CUL 100/0,15) which are further thermalized by heatsinks mounted on the MXC before connecting to a custom 40-pin sample socket (see sample holder section below). The total resistance of one wire from the room temperature BNC termination to the millikelvin sample stage is ~ 3 kΩ. The combination of Thermocoax cables and the low-temperature RC filters attenuates the high-frequency noises that would otherwise heat the sample. A similar filtering setup has previously been used to achieve an electron temperature down to 12 mK.[15]

### D. Sample Holder

The sample holder is composed of a custom 40-pin socket with a large central hole to allow FIR/THz light to pass through the sample to be absorbed by the surrounding shielding of the 1K stage or to be detected via a detector located below the sample to enable transmission measurement mode (not yet enabled in the current setup). The design and an optical image of the holder are shown in Fig. 3. The holder is created using an FR-4 plate with press-fit gold-plated brass pin receptacles. The holder is mounted to a gold-coated copper plate and supported by four gold-coated copper rods anchored to the MXC plate. Two copper braids are mounted to the plate to provide additional cooling power.

The chip carrier is made of FR-4, coated with silver for better thermal conductivity. 40 gold-plated brass pins, matched to the receptacles, are wire bonded to the samples. The entire sample mounting system, the optical head, and the still shield of the DR are fitted into the magnet bore with a diameter of 150 mm. The sample is located at the center of the magnet. The superconducting magnet used in the current setup can reach 5 T.

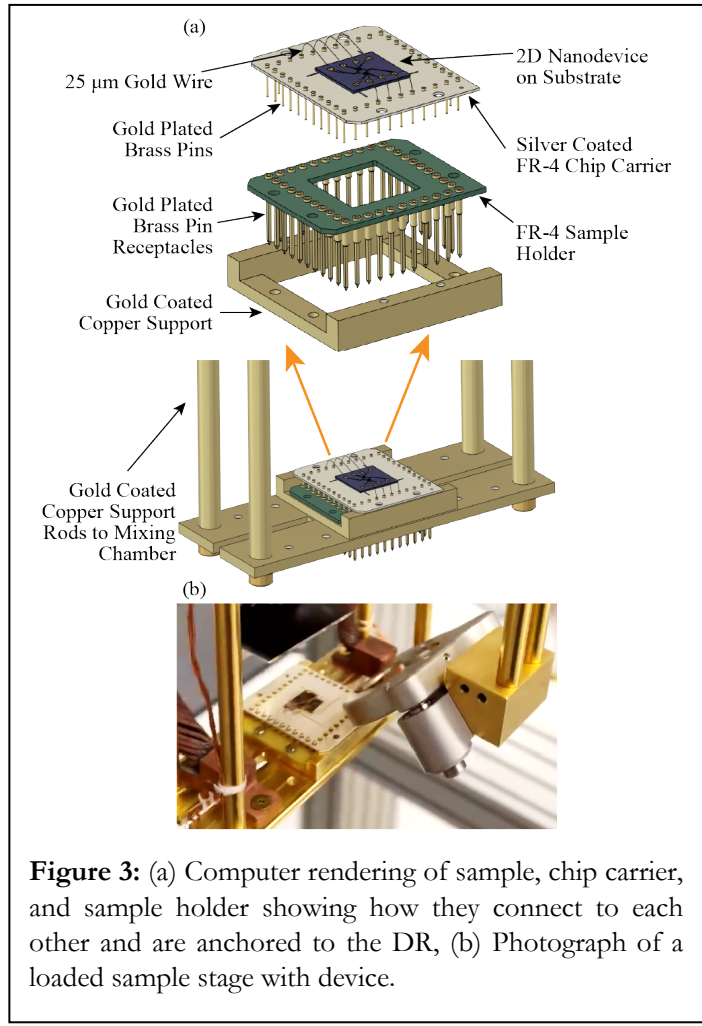

**Figure 3:** (a) Computer rendering of sample, chip carrier, and sample holder showing how they connect to each other and are anchored to the DR, (b) Photograph of a loaded sample stage with device.

## III. Evaluation

We next characterize the temperatures of both the MXC plate and the sample, which are the essential features of our setup.

### A. MXC Temperature

The DR, as delivered, can reach a base temperature as low as ~ 7 mK before the installation of the optical components. The fridge has a cooling power of ~ 500 μW at 100 mK. The addition of the optical window, FIR/THz optics, filters, and mechanical supports introduced multiple new heat loads as well as new thermal connections between various cooling stages within the DR. We characterize the performance of the MXC temperature under various configurations in Fig. 4. One important thermal source is from the optical window, which allows for not only the FIR/THz light but also blackbody radiation from room temperature objects to enter the fridge (Fig. 4a). The blackbody spectrum of a 300 K object is shown in Fig. 4b, which covers the FIR/THz range and must be carefully managed when performing experiments in this regime. Fig. 4c plots the MXC temperature ($T_{MXC}$) during the condensation of the helium mixture with three different open apertures on the three inner cans (Fig. 4a), which are thermalized to 50 K, 4 K, and 1 K respectively. Without any low-temperature optical filters in place (i.e., all blackbody radiation comes through), $T_{MXC}$ reaches 246 mK (for 25 mm open apertures), 190 mK (for 15 mm), and 72 mK (for 9 mm).

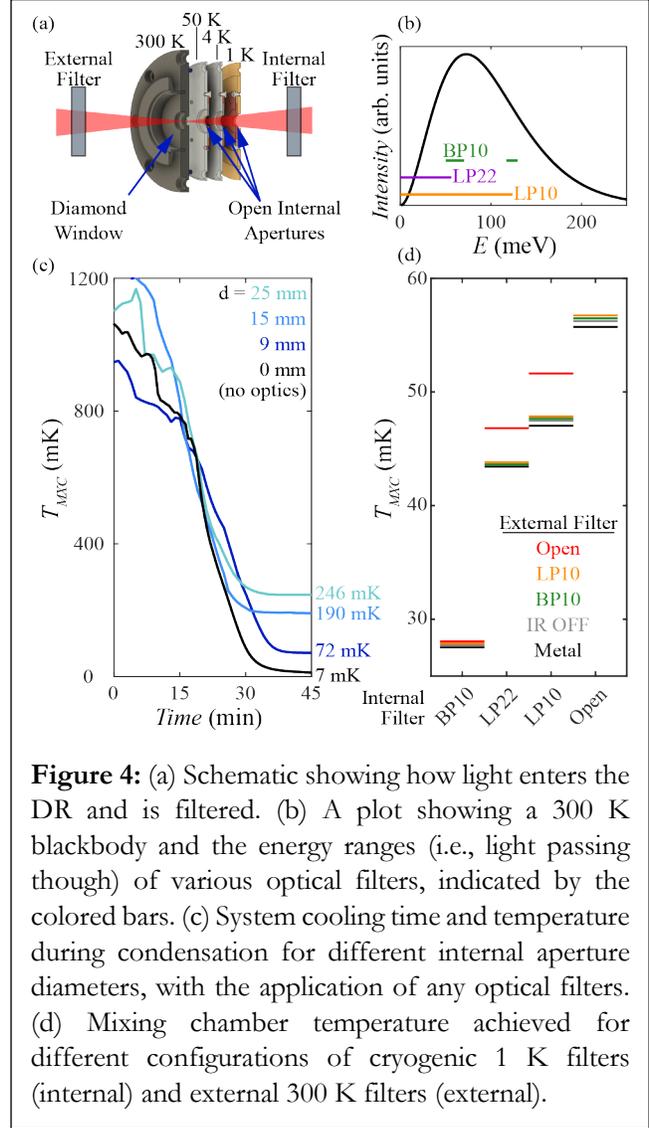

**Figure 4:** (a) Schematic showing how light enters the DR and is filtered. (b) A plot showing a 300 K blackbody and the energy ranges (i.e., light passing though) of various optical filters, indicated by the colored bars. (c) System cooling time and temperature during condensation for different internal aperture diameters, with the application of any optical filters. (d) Mixing chamber temperature achieved for different configurations of cryogenic 1 K filters (internal) and external 300 K filters (external).

In experiments, we typically choose a 9 mm open aperture to achieve the lowest temperature, which will be used for all subsequent measurements shown. To further reduce the blackbody radiation, we install long-pass (LP) (Infrared Laboratories, OF-P10-C & OF-P22-C) or band-pass (BP) (Thorlabs FB10000-500) optical filters on the low-temperature optical filter stage (as described in section II-B). Fig. 4d plots $T_{MXC}$ after mixture condensation under various combinations of internal and external filters, with fixed 9 mm open apertures on the inner stages. The transmission range of the different filters is sketched in Fig. 4b. We clearly see that the filter has a strong impact on the $T_{MXC}$. With an internal long pass filter of LP22 (< 55 meV pass), we achieve $T_{MXC}$ ~ 44 mK, a remarkable value that enables millikelvin spectroscopic studies at this FIR/THz range, which is relevant for many interesting quantum phenomena in correlated materials. We also note that the FIR/THz source power needed for the photocurrent FTIR spectroscopy, to be elaborated below, is low (typically ~ 5 μW on the device), and hence does not substantially affect $T_{MXC}$.

## B. Sample Electron Temperature

We characterize the sample electron temperature, $T_e$, using a monolayer $WTe_2$ device fabricated on an insulating $Si/SiO_2$ substrate as a thermometer. The detailed fabrication and characterization of the device can be found in our previous work[16,17]. Monolayer $WTe_2$ is an excitonic quantum spin Hall insulator[17–19], which can be converted to a superconductor at ULT when it is electrostatically gated[20,21]. The resistance map, reflecting its electronic phase diagram, of the monolayer device is shown in Fig. 5a. Near the superconducting transition at the high electron doping ($n_g$), its resistance is highly sensitive to electron temperature and hence can be used for calibration of $T_e$. Fig. 5b plots the temperature-dependent four-probe resistance $R_{xx}$ of the device at selected doping in the superconducting region (a small magnetic field is applied to broaden the transition for better calibration), when the radiation on the sample is minimized (brown reference curve). Namely, the radiation focus is moved off of the sample via the attocube positioner and the external and internal filters are replaced with pieces of aluminum to block any radiation. To calibrate the effect of applying an FIR/THz beam, we then measure the same temperature-dependent $R_{xx}$ under various combinations of external and internal filters light focused onto the sample (gray, green, and red curves). The curves taken under radiation (Fig. 5b) show higher saturated resistances, signifying an increased $T_e$. We refer to $T_e$ as the temperature where the resistance of the brown reference curve matches the saturated resistance for various excitation conditions, as shown by the dashed lines in Fig. 5b. The extracted $T_e$ is plotted in Fig. 5c.

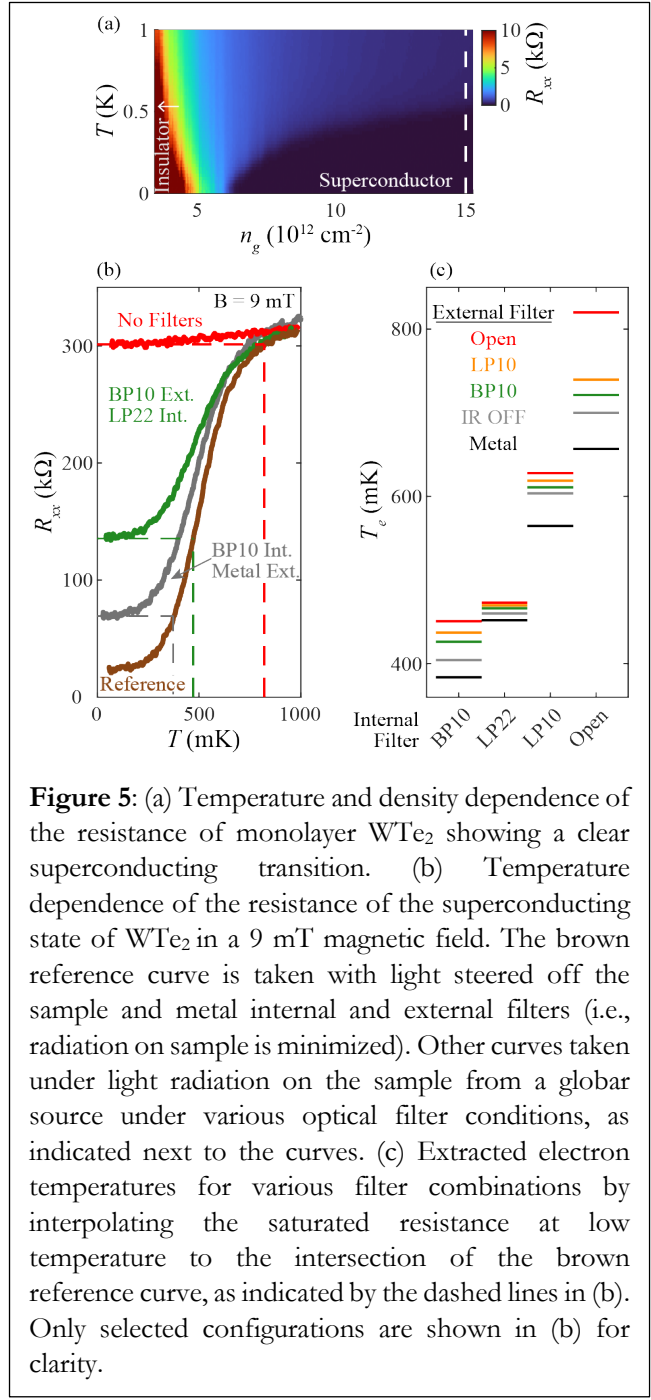

**Figure 5**: (a) Temperature and density dependence of the resistance of monolayer $WTe_2$ showing a clear superconducting transition. (b) Temperature dependence of the resistance of the superconducting state of $WTe_2$ in a 9 mT magnetic field. The brown reference curve is taken with light steered off the sample and metal internal and external filters (i.e., radiation on sample is minimized). Other curves taken under light radiation on the sample from a globar source under various optical filter conditions, as indicated next to the curves. (c) Extracted electron temperatures for various filter combinations by interpolating the saturated resistance at low temperature to the intersection of the brown reference curve, as indicated by the dashed lines in (b). Only selected configurations are shown in (b) for clarity.

The relative values and trend are consistent with the $T_{MXC}$ plotted in Fig. 4d, yet $T_e$ is substantially higher than $T_{MXC}$. This is expected as all radiation is focused on the sample area. Remarkably, we still consistently obtain sub-kelvin electron temperatures in our setup, and with the optimized combination of filters a $T_e$ of ~ 450 mK under FIR/THz radiation is achieved.

We comment on several aspects. First, here we use a standard 2D device based on monolayer $WTe_2$, consisting of typical materials (silicon substrate, boron-nitride, metal electrodes, and graphite gates), for calibrating the $T_e$. This is critical for demonstrating the performance of our setup for probing 2D devices since device details, such as composition materials and thickness, etc., are expected to impact infrared absorption and hence the electronic temperature. The same device will be used for implementing spectroscopic measurement, as elaborated in the next section. Second, thermal radiation from high-temperature objects is the main source that heats the sample in our measurement scheme. As a result, the application of the internal optical filter has a strong effect on the sample temperature. Third, under certain filter settings, a substantial amount of thermal radiation still reaches the sample. For example, the long-pass filter LP22 only cuts off photons at energies higher than 55 meV (Fig. 4b). For future applications, if the target spectral range is at much lower energy, e.g., < 10 meV in the THz regime, then one can apply a different optical filter that cuts off photons at even lower energies, which will future reduce $T_e$. In this regime, we believe $T_e$ < 300 mK is reachable. Fourth, our characterization device is fabricated on a Si substrate, which is known to absorb FIR/THz light. Replacing the Si substrate with an IR transparent substrate will help further reduce $T_e$.

### IV. Implementing FTIR Spectroscopy

We next demonstrate the implementation of photoconductivity-based FTIR spectroscopy on micron-sized samples at ULT, first on monolayer $WTe_2$ and then on a thick $1T-TaS_2$ sample.

#### A. Spectroscopy using a superconducting monolayer as a detector

In the previous section, we described the use of a monolayer $WTe_2$ device for calibrating $T_e$. The same sample also performs as a photodetector and can be used for conducting FTIR spectroscopy. The simplified measurement scheme and the optical images of the devices used are shown in Fig. 6a & b. The beam is modulated by a mechanical chopper (typically at a frequency of ~190 Hz) to enable sensitive photocurrent or photovoltage measurements of the sample using lock-in techniques, as typically used in transport measurements. The light has been filtered using a combination of low-temperature and room-temperature filters to produce a peak of around 55 meV with a distinctive shape.

The grey curve in Fig. 6c plots the differential resistance of the monolayer $WTe_2$ superconductor (electron-doped), without applying IR radiation, as a function of a DC current bias ($I_{dc}$) applied to the source electrode, showing the typical nonlinear characteristic of a superconductor. The black curve is the same differential resistance measurement but with IR radiation, i.e., the beam is focused on the sample after passing an external band pass filter (BP10) and an internal low pass filter (LP22) (see Fig. 4b for the working energy ranges of the filters.). It is clear that the superconducting state still survives under radiation, again proving the ultralow $T_e$ since the superconducting critical temperature is around 480 mK. The photovoltage ($V_{PC}$) between two voltage probes measured at the same condition (source biased and drain grounded, see Fig. 6a) is shown as the blue curve, which displays an interesting enhancement near the critical current (indicated by the vertical arrows in Fig. 6c). We set $I_{dc}$ at the value where $V_{PC}$ is maximized and then perform FTIR spectroscopy by moving the mirror to alter the delay time between the two interfering beams. The resulting interferogram is plotted as the inset in Fig. 6d, where a clear oscillating pattern is resolved. A fast Fourier transformation (FFT) leads to a spectrum shown in Fig. 6d, which indeed reproduces the filtered source that we have directed onto the sample. Note that in this measurement, WTe$_2$ is used simply as a ULT photodetector, and the observed spectral features are not from WTe$_2$ but from the filtered light source. With a wider optical window and other improvements, one might see intrinsic features of WTe$_2$ itself. Here we only demonstrate that our instrument is functioning as expected and leave a careful investigation of WTe$_2$ for future studies.

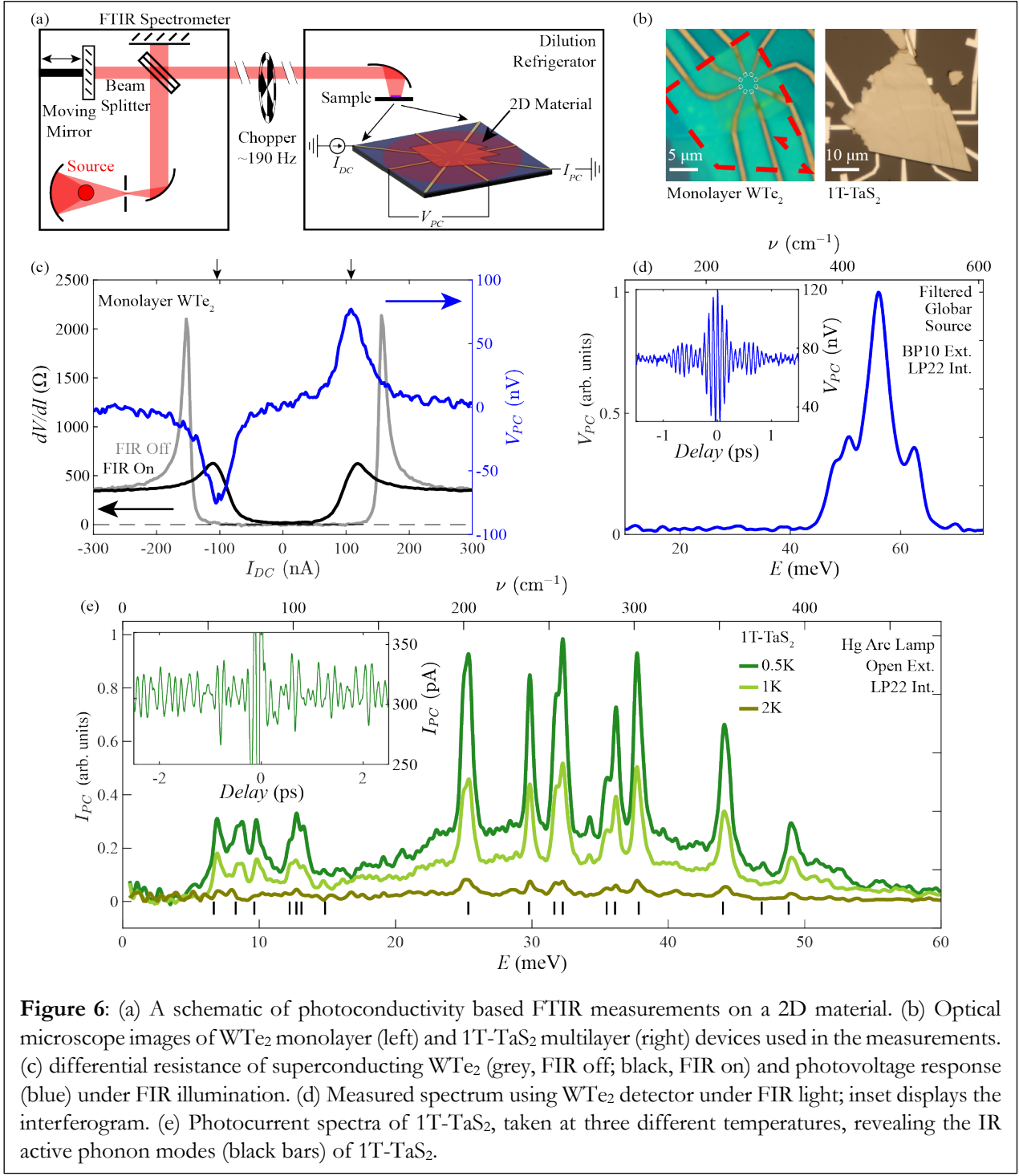

**Figure 6**: (a) A schematic of photoconductivity based FTIR measurements on a 2D material. (b) Optical microscope images of WTe$_2$ monolayer (left) and 1T-TaS$_2$ multilayer (right) devices used in the measurements. (c) differential resistance of superconducting WTe$_2$ (grey, FIR off; black, FIR on) and photovoltage response (blue) under FIR illumination. (d) Measured spectrum using WTe$_2$ detector under FIR light; inset displays the interferogram. (e) Photocurrent spectra of 1T-TaS$_2$, taken at three different temperatures, revealing the IR active phonon modes (black bars) of 1T-TaS$_2$.

### B. Detection of low energy excitations of insulating 1T-TaS$_2$

In addition to work on superconducting materials, here we also demonstrate that this FTIR scheme also works well for insulating samples, using 1T-TaS$_2$, a material that is known to host interesting charge density wave (CDW) phases[22–24] and is proposed as a quantum spin liquid candidate[25,26]. We fabricate a small-size multilayer 1T-TaS$_2$ flake with electrodes (Fig. 6b). At ULT, the resistance of this insulating device is large and shows a strong temperature-dependent response. 1T-TaS$_2$ hosts multiple IR-active phonon modes, associated with its CDW phase at low temperatures. Previous infrared measurements on this material have only been conducted on large bulk crystals, but not nanofabricated small samples. Here we demonstrate FTIR spectrometry on small, exfoliated samples using our instrument, which paves the way for optical investigation of this or similar materials down toward their monolayer limits.

Similar to the measurements on the WTe$_2$ device, here we record photocurrent generated on 1T-TaS$_2$ as a function of the FTIR delay time. A typical interferogram is shown in Fig. 6e inset, an FFT of which produces multiple low energy peaks below 50 meV (Fig. 6e). They agree very well with previous IR measurements on bulk crystals[27], and they have been identified as IR-active phonon modes of 1T-TaS$_2$. The optical charge gap is around ~ 100 meV in this material[27]. The results demonstrate that we have successfully detected charge-neutral modes well below the charge gap at ULT. The spectrum taken at higher temperatures is also shown in Fig. 6e, but the signal quickly becomes undetectable. This is because at ULT, sample resistance is very sensitive to temperature and hence the FIR/THz radiation, which enables 1T-TaS$_2$ to act as a good detector. Previous FIR/THz measurements revealing the same modes were conducted using reflection or transmission measurements (which require large samples) at high temperatures (> 10 K)[27]. Our experiment enables FIR/THz detection of micron-sized samples at ULT, which may be extended to monolayers for examining interesting neutral electronic modes. We leave such investigations for future study.

### V.     **Discussion & Outlook:**

In this work, we report the design and creation of a versatile optical and optoelectronic platform for investigating quantum materials at ULT, integrating both optical and electronic transport measurements. We demonstrate photoconductivity-based FTIR spectroscopy at electron temperatures as low as ~ 450 mK and performed measurements on micron-sized samples including both WTe$_2$ monolayer and 1T-TaS$_2$ multilayers, paving the way for further investigating 2D materials and van der Waals heterostructures. Near-term improvements include further reducing the electron temperature with better thermal engineering, increasing the magnetic field strength, and expanding to an even lower spectral range of sub-meV. Additionally, experiments examining polarization-dependent responses may be enabled by mounting polarizers and quarter waveplates. With future modifications, a variety of interesting optical measurements may be enabled in this all-free space platform, including potentially near-field nanoscale microscopy if an atomic force microscope tip is engaged. Simultaneous detection of excitation in both energy and spatial domains would promote the power of the instrument to the next level. Overall, given the advances in 2D materials and other novel condensed matter systems, which have proven their richness as sources of new quantum matter, the platform developed here is promising to enable new discoveries of fundamental quantum phenomena.

## Acknowledgments

We acknowledge support from Gordon and Betty Moore Foundation's EPiQS Initiative, Grant GBMF11946 to S.W. This work is also supported by ONR through a Young Investigator Award (N00014-21-1-2804), AFOSR through a Young Investigator Program award (FA9550-23-1-0140), the Materials Research Science and Engineering Center (MRSEC) program of the National Science Foundation (DMR-2011750) and the Eric and Wendy Schmidt Transformative Technology Fund at Princeton. Device fabrications are supported by NSF through a CAREER award (DMR-1942942) to S.W. L.M.S. acknowledges support from the Gordon and Betty Moore Foundation through Grants GBMF9064 and the David and Lucile Packard Foundation and the Sloan Foundation. Y.J. acknowledges support from the Princeton Charlotte Elizabeth Procter Fellowship program. T.S. acknowledges support from the Princeton Physics Dicke Fellowship program. A.J.U acknowledges support from the Rothschild Foundation and the Zuckerman Foundation. J.F.K. acknowledges support from the Arnold O. Beckman Postdoctoral Fellowship.

## Author contributions

S.W. conceived and supervised the project. A.J.U., M.O., P.W. and S.W. designed the instrument. M.O. and A.J.U. built the optical, mechanical, vacuum and software systems. Y.T., P.W. and Y.J. built electronic wiring and transport measurement systems. M.O. and A.J.U. performed measurements and characterization and analyzed data, assisted by P.W., Y.T., G.Y., Y.J., and T.S. R.S., J.F.K., and L.M.S. grew bulk $WTe_2$ and $1T\text{-}TaS_2$ crystals. A.J.U and M.O. fabricated the $1T\text{-}TaS_2$ device. G.Y. and P.W. fabricated the $WTe_2$ device. S.W., M.O. and A.J.U. wrote the paper with input from all authors.

## Competing interests

Authors declare that they have no competing interests.

## Data availability

All data needed to evaluate the conclusions in the paper are present in the paper. Additional data related to this paper are available from the corresponding author upon reasonable request.